\documentclass[sigconf]{acmart}

\usepackage{amsmath}

\AtBeginDocument{%
  \providecommand\BibTeX{{%
    \normalfont B\kern-0.5em{\scshape i\kern-0.25em b}\kern-0.8em\TeX}}}

\copyrightyear{2023}
\acmYear{2023}
\setcopyright{acmlicensed}
\acmConference[WWW '23 Companion]{Companion Proceedings of the ACM Web Conference 2023}{April 30-May 4, 2023}{Austin, TX, USA}
\acmBooktitle{Companion Proceedings of the ACM Web Conference 2023 (WWW '23 Companion), April 30-May 4, 2023, Austin, TX, USA}
\acmPrice{15.00}
\acmDOI{10.1145/3543873.3587569}
\acmISBN{978-1-4503-9419-2/23/04}

\begin{document}

\title{Online to Offline Crossover of White Supremacist Propaganda}

\author{Ahmad Diab}
\affiliation{
  \institution{University of Pittsburgh}
   \city{Pittsburgh}
   \country{USA}}
 \email{ahd23@pitt.edu}

 \author{Bolor-Erdene Jagdagdorj}
 \affiliation{%
   \institution{Carnegie Mellon University}
   \city{Pittsburgh}
   \country{USA}}
 \email{bjagdagd@andrew.cmu.edu}

 \author{Lynnette Hui Xian Ng}
 \affiliation{%
   \institution{Carnegie Mellon University}
   \city{Pittsburgh}
   \country{USA}}
 \email{lynnetteng@cmu.edu}

 \author{Yu-Ru Lin}
 \affiliation{%
   \institution{University of Pittsburgh}
   \city{Pittsburgh}
   \country{USA}}
 \email{yurulin@pitt.edu}

 \author{Michael Miller Yoder}
 \affiliation{%
   \institution{Carnegie Mellon University}
   \city{Pittsburgh}
   \country{USA}}
 \email{yoder@cs.cmu.edu}

\renewcommand{\shortauthors}{Diab, et al.}

\begin{abstract}
White supremacist extremist groups are a significant domestic terror threat in many Western nations.
These groups harness the Internet to spread their ideology via online platforms: blogs, chat rooms, forums, and social media, which can inspire violence offline.
In this work, we study the persistence and reach of white supremacist propaganda in both online and offline environments.
We also study patterns in narratives that crossover from online to offline environments, or vice versa. 
From a geospatial analysis, we find that offline propaganda is geographically widespread in the United States, with a slight tendency toward Northeastern states.
Propaganda that spreads the farthest and lasts the longest has a patriotic framing and is short, memorable, and repeatable.
Through text comparison methods, we illustrate that online propaganda typically leads the appearance of the same propaganda in offline flyers, banners, and graffiti. 
We hope that this study sheds light on the characteristics of persistent white supremacist narratives both online and offline.
\end{abstract}

\ccsdesc[500]{Human-centered computing~Collaborative and social computing}
\ccsdesc[300]{Social and professional topics~User characteristics}
\ccsdesc[300]{Information systems~Social networking sites}


\keywords{white supremacist, hate speech, propaganda, geospatial, temporal, narratives}

\maketitle

\section{Introduction}
In 2021, top officials from the US Department of Justice and Department of Homeland Security named white supremacist extremism as the most significant domestic terror threat~\citep{sullivan_top_2021}.
This ideology has inspired extremists to commit mass murders in Pittsburgh \cite{phillips2018daily}, El Paso \cite{quek2019paso}, and Buffalo \cite{somoano2022lessons} in recent years.
Far-right groups with white supremacist and fascist elements have recently led or planned violent attacks against governments in the United States, Brazil, and Germany.
To spread their ideology and build communities of hate, white supremacist groups disseminate propaganda using memorable slogans like ``end all immigration'' in both the online and offline spaces \cite{keum2022hate}. 
With this backdrop, it is increasingly important to study the spread of hate by white supremacist groups in the online medium, and how it relates to offline propaganda.

Online hate speech is a frequent topic of attention in the cyber social threat space \cite{fortuna2018survey,chetty2018hate,alorainy_enemy_2019}.
Some quantitative and computational work has focused specifically on white supremacist ideology. 
\citet{alatawi_detecting_2021} develop methods for detecting white supremacist hate speech, while \citet{eddington2018communicative} draw network connections between white supremacy hate groups within the United States and United Kingdom in 2016.
However, fewer studies examine the connections between offline events and online hate. 
\citet{lupu2023offline} observed that offline trigger events, such as protests and events, often leads to increase in online hate speech; \citet{hirvonen2013sweden} studied the affinity between online white supremacy hate speech and offline hostility; and \citet{keum2022hate} identified that the constant dissemination of ideology by white supremacist groups online can impact the mental health of their surrounding community. Given these potential dangers, it is important to characterize and study the patterns of messaging that draws people into white supremacism and investigate how this messaging connects to offline violence \cite{gagliardone2015countering}.

This paper aims to improve our understanding of the correlation between online and offline white supremacist messaging. 
Using an overlay of offline and online sources, we identify matching white supremacist propaganda and investigate the following research questions:
\begin{quote}
    \textbf{RQ1}: What spatial and temporal patterns characterize the spread of offline white supremacist propaganda? Which types of propaganda spread the furthest and last the longest? 
\end{quote}
Through a spatial and geographic analysis of events reported as containing offline white supremacist propaganda (flyers, graffiti, banners, etc), we identify when the same propaganda is used across events and analyze the reach of each propaganda phrase.

\begin{quote}
    \item[] \textbf{RQ2}: What temporal patterns characterize the online-to-offline crossover of white supremacist propaganda? 
\end{quote}
We directly correlate online and offline white supremacist ``quotes'', which are phrases representing ideologies and propaganda promoted by individuals and extracted from online posts and offline event descriptions. We do so via text similarity matches, serving as indicators to identify the persistence of the quotes in both spaces.
Similar methods to identify persistence of text across multiple platforms were conducted in studies of the spread of news \cite{leskovec2009meme} and disinformation \cite{ng2020analyzing}. 
Using information on quote persistence, we characterize the geographical, time-series and narrative distributions of quotes that persist in both online and offline environments.

\subsection{Contributions}
To our knowledge, this study is the first of its kind to combine the analysis of white supremacist propaganda across both offline and online environments, studying a US-wide reach of white supremacist ideology. 
Using a combination of geographical, temporal and text analysis, we make the following contributions:
\begin{enumerate}
    \item We measure the persistence of offline white supremacist propaganda messages through text analysis techniques and quantitatively estimate the spatiotemporal distribution of particular white supremacy slogans.
    \item We analyze the prevailing common narrative themes between online and offline white supremacist propaganda.
    \item We find preliminary evidence within 2 years of overlapping data that white supremacist propaganda quotes begin online before appearing in offline events, supporting the hypothesis that online activity can translate into offline activity. 
\end{enumerate}

\section{Related Work}

\subsection{White Supremacist Extremism}
Merriam-Webster defines white supremacy as ``the belief that the white race is inherently superior to other races and that white people should have control over people of other races''\footnote{\url{https://www.merriam-webster.com/dictionary/white\%20supremacy}, accessed 24 January 2023}. Social movements organized under this principle are recognized as white supremacist extremism. 
Key beliefs of this ideology include essentialist, ``natural'' race and gender hierarchies with white men at the top \cite{brown_wwwhatecom_2009}. 
White supremacists believe that white male power is threatened in today's world and that violent action is necessary to protect the white race \cite{ansah_violent_2021,ferber_reading_2000}.
White supremacist extremism has been demonstrated in premeditated violent attacks against people of other races, which provides white supremacists a shared sense of accomplishment, solidarity and power \cite{windisch2018understanding}. 
Larger scale incidents include the Tree of Life synagogue attack in Pittsburgh, Pennsylvania, where the shooter believed Western Civilization was facing ``extinction'' and that Jewish refugees were ``invaders''; the 2017 ``Unite the Right'' rally in Charlottesville, North Carolina, where an anti-racist protester was killed with a car~\cite{blout2020white}; anti-Muslim shootings in Christchurch, New Zealand; and the anti-Latinx shooting in El Paso, Texas inspired by the Christchurch event.

Though these movements are regarded as fringe, scholars such as sociologist Jessie Daniels \citep{daniels2009cyber} do not view them as isolated from structural white supremacism, a system in which white people control the material resources and major institutions of societies~\citep{ansley_stirring_1989,caswell_urgent_2021}.
These social movements exploit bigotries widely held in societies with structural white supremacism~\citep{ferber2004home,berlet_overview_2006,pruden_birds_2022}.
Ideologies that are implicit in structural white supremacy are explicitly stated in white supremacist social movements. 
Studying white supremacist communication is thus important not just to understand a fringe social movement, but also as a window into broader societies organized under white supremacy.

Offline white supremacist propaganda is a small piece of the larger communication infrastructure of white supremacist extremism. To better understand white supremacist extremism, we must also look online. White supremacist groups have recognized the potential of the Internet in spreading their message. Donald Black, the founder of white supremacist forum Stormfront, encouraged his followers to use social networking sites to ``reach new people and bring them here'' \cite{perry2016white}. Similarly, David Duke, former Ku Klux Klan leader, mentioned in 2007 that the Internet could give ``millions access to the truth [...] of racial enlightenment'' \cite{daniels2009cyber}. 

White supremacist groups have made extensive use of social media and online forums for information provision, seeking donations, recruitment and networking \cite{wong2015supremacy}. They have also united around political campaigns, discursively creating and connecting their organizations and lingo to the campaigns \cite{eddington2018communicative}, and made use of online chat channels to distribute and mobilize extremist movements \cite{guhl2020safe}.

Therefore, it is important to trace the presence of the same white supremacist propaganda in both online and offline environments.

\subsection{The Intersection between Online and Offline Speech}
The spread of online propaganda and offline activism has been observed to be positively related. Online platforms facilitate the spread of white supremacist ideology through community building and the development of shared realities and emotions. Empirical histories suggest that the sowing of discord among individual and online communities can create an impetus for violent offline expressions, such as protests and shootings \cite{greijdanus2020psychology}. 

The formation of online communities is rarely isolated: online communities signal a social connection in the offline world. Literature provides a mixed view of the connection between the online to offline ecology: a supporting relationship and a negative relationship. Spread of ideologies online can be dismissed as unproductive and inhibit the incidents of offline activism \cite{wilkins2019all}. For example, activists online have been observed to distance themselves from offline riots, contrasting away the online-offline domains \cite{lefebvre2018grievance}. However, online efforts such as the production of videoclips can crossover to offline action, where these videoclips inspire extremism or teach the ways of extremism (e.g., how to fire a gun) \cite{cohen2014detecting}.
Others find that both online speech and offline events can reinforce each other in a cycle of radicalization \cite{gallacher_mutual_2021,hutchinson_online_2022}.

Given the close relationship and the dynamic interplay between the online and offline domains, it is clear that one domain influences the other, and vice versa. As such, it is important to understand the intersection between online and offline domains.

\subsection{Computational Propaganda}
Computational propaganda is the ``use of algorithms, automation and human curation to purposefully distribute misleading information'' \cite{woolley2017computational}. The use of computational methods increases the ease and rate of disseminating propaganda information, and provides flexibility for both broad efforts and targeted attempts at distributing propaganda materials. The presence of computational propaganda has been observed most prominently in the political sphere, where automated agents have attempted to influence the UK-EU Brexit referendum \cite{howard2016bots}, the US elections \cite{woolley2017computational} and the Brazil elections \cite{arnaudo2017computational}, to name a few instances.

In recent years, detection of computational propaganda has made significant progress. Several shared tasks involving detection and classification of annotated propaganda datasets have been constructed for the development of detection and analysis techniques within the community \cite{10.1145/3488560.3501395,shaar-etal-2021-findings}. Detection methods involve constructing machine learning models for identification of the use of propaganda techniques \cite{da-san-martino-etal-2019-fine} and the classification of the types of techniques employed \cite{barron2019proppy} within texts, and the use of network analysis approaches to identify propaganda through the presence of coordinated inauthentic action \cite{da2021survey}.

Our work focuses on how online white supremacist propaganda of all types also appears in offline propaganda.


\section{Datasets}
In this study, we use two datasets that contain white supremacist propaganda: one offline dataset and one online dataset. We elaborate on the details of these datasets in the following section.

\subsection{Offline Dataset}
We gathered 43,154 events involving white supremacist propaganda from the ADL H.E.A.T. Map\footnote{\url{https://www.adl.org/resources/tools-to-track-hate/heat-map}}. 
Collected by the Anti-Defamation League, an anti-hate non-profit organization, this map aggregates and plots incidents of hate, extremism, antisemitism and terrorism from a variety of sources, including news reports, government documents and victim reports. 
Commonly, events include propaganda such as flyers and banners, with a description of the incident and metadata including location, date, group, and ideology of the group.
The data fields that are extracted from the events are reflected in \autoref{tab:offline_dataset}. 
The data considered in this study spans from January 2008, until July 2022.

\begin{table}
\centering
\begin{tabular}{| l | l |}
\hline
\textbf{Data Field} & \textbf{Description} \\ 
\hline
quote & white supremacist propaganda text\\ 
location & city and state of event \\
event & event details (i.e. people/organizations involved) \\
timestamp & time/date of event occurrence \\
\hline
\end{tabular}
\caption{Data fields extracted per event for offline dataset }
\label{tab:offline_dataset}
\end{table}

\paragraph{Location Annotation} For each offline white supremacist propaganda phrase, or ``quote'', we annotated it with latitude/longitude (lat/lon) coordinates by parsing the provided location information (city, state) through the Nominatim API\footnote{\url{https://nominatim.org/}}. The API takes in a (city, state) pair and returns the (lat/lon) coordinates to which the pair is geographically located. Errors in the output due to misspelled cities (e.g. Mattson, IL instead of Matteson, IL) and ambiguous locations (e.g. national park addresses) were manually corrected.

\subsection{Online Dataset}
The online dataset contains 4,371,453 posts after removing duplicates, it spans from October 2001 through November 2019. The data was collected from explicitly white supremacist domains and organizations, assembled from a variety of datasets and data dumps by \cite{anonymous2023}. 
\autoref{tab:online_dataset} lists the data fields extracted per post from the online dataset. Further details about the online dataset are reflected in \autoref{tab:ws_data_full}.

\begin{table}
\centering
\begin{tabular}{| l | l |}
\hline
\textbf{Data Field} & \textbf{Description} \\ 
\hline
quote & white supremacist propaganda text\\ 
dataset & data source of posts (cf. \autoref{tab:ws_data_full})\\
platform & online platform source (cf. \autoref{tab:ws_data_full})  \\
timestamp & time/date of event occurrence \\
\hline
\end{tabular}
\caption{Data fields extracted per post in online dataset}
\label{tab:online_dataset}
\end{table}


\begin{table*}[tb]
\begin{tabular}{| l | l | r | r |}
\hline    
Platform & Data source & \# Posts & \# matched offline quotes\\
\hline    
4chan & \citet{papasavva_raiders_2020} & 2,686,267 & 137,744\\
4chan & \citet{jokubauskaite_generally_2020} & 578,650  & 20,546\\
4chan, 8chan, Reddit & \citet{rieger_assessing_2021} & 361 & 65\\
Stormfront & Stormfront dump & 751,980  & 57,761\\
Iron March & Iron March dump & 179,468  & 10,559\\
Twitter & \citet{qian_hierarchical_2018} & 84,695 & 2409\\
Twitter & \citet{elsherief-etal-2021-latent} & 3,480 & 261\\
Twitter & \citet{alatawi_detecting_2021} & 1,098 & 46\\
Twitter & \citet{siegel_trumping_2021} & 170 & 37\\
Discord & Patriot Front dump & 39,577 & 675\\
Daily Stormer & \citet{calderon_linguistic_2021} & 26,099 & 5058 \\
\hline
\end{tabular}
\caption{Details of online dataset.}
\label{tab:ws_data_full}
\end{table*}

This dataset includes forums and websites dedicated to white supremacism, including Stormfront, Iron March, and the Daily Stormer.
Tweets from organizations that the Southern Poverty Law Center labels as white supremacist hate groups~\citep{qian_hierarchical_2018,elsherief-etal-2021-latent} are also included.
This dataset filters 4chan /pol/ data, an imageboard known for white supremacy, to posts from users choosing Nazi, Confederate, Fascist, and White Supremacist flags (in the dataset from \citet{papasavva_raiders_2020}), as well as posts in ``general'' threads with fascist and white supremacist topics (in the dataset from \citet{jokubauskaite_generally_2020}).
We also include smaller datasets manually annotated for white supremacist ideology from \citet{rieger_assessing_2021}, \citet{siegel_trumping_2021}, and \citet{alatawi_detecting_2021}.

\section{Methods}

\subsection{Matching Propaganda Quotes}
\label{sec:matching}
For this work, we refer to individual propaganda slogans or phrases used by individuals or organizations as ``quotes''. Through this paper, we report these quotes within double quotation marks (`` ''). 
For example, an incident where Patriot Front, a white supremacist group, distributed propaganda with the slogan ``America is not for sale'' and the incident where an unknown person drew the slogan ``America is not for sale'' in graffiti are incidents that share the same propaganda, though they differ in nature, location and people involved. Mapping out such connections can aid in identifying the spatial and temporal spread of similar white supremacist propaganda.
Identifying quotes offline and locating them online helps establish connections between the same propaganda ideology spreading in both domains.
We detail our process for matching quotes in this section.

We start with the set $Q$ of all offline quote occurrences, extracted from events in the offline dataset.
For each pair of quote occurrences $q_1, q_2 \in Q$, we perform two steps of matching comparison to determine whether they are similar. 
The first step simply compares if the texts of the two quotes (lowercased and tokenized) are exactly the same. 
Should the first step return negative, the quotes are compared in a second step. 
The second step further pre-processes both quotes by removing punctuation and English stopwords. 
The pre-processed quotes are then compared with each other again; if they match exactly, they are considered as matches.

This matching flow was performed in three iterations: (1) offline-offline quotes, to identify matching quote clusters in the offline dataset; (2) offline-online, to identify quotes that had crossed over and appeared in both medium; and (3) online-online quotes, to identify matching quote clusters in the online dataset. In each iteration, we performed an all-pairs matching flow, meaning every quote was compared to all other quotes in the dataset. 
After each matching iteration, the generated unique set of quotes were manually inspected and combined if necessary. 

In the offline dataset, we identified 1798 unique white supremacist propaganda quotes.
Not all reported quotes are what would generally be regarded as propaganda, however.
Some quotes are very short (e.g. ``aryan'') or non-ideological (e.g. ``send a message'').
In order to filter these out, two of the authors annotated quotes for being ideological, not just descriptive.
Specifically, annotators marked whether each phrase portrays a white supremacist ideological position or evaluation (such as ``race traitors'') instead of being descriptive and likely also used by those without a white supremacist worldview (such as the phrase ``white nationalists'').
Annotators discussed difficult cases and came to a consensus.
1655 out of 1798 unique quotes (92.0\%) were annotated as propaganda.

For the online dataset, we found 188,861 online posts containing exact matches of offline quotes.
Searching for near matches (removing punctuation and stopwords) provided an additional 46,701 posts, yielding a total of 235,562 online posts that mention either the exact or near matches of propaganda quotes in offline events.

\subsection{Spatial Coverage of Offline Propaganda Quotes}
\begin{figure}[h!]
\centering
\includegraphics[width=0.5\textwidth]{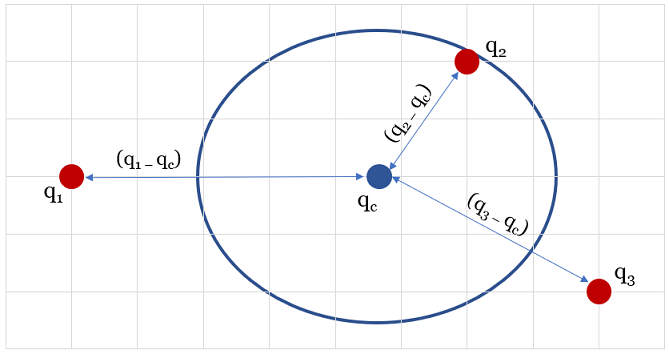}
\caption{Illustration of parameters used in calculating the radius of gyration for $N=3$ matching quotes $q_1$, $q_2$, and $q_3$. The blue circle is the geographical cover (gyration) of the quote.}
\label{fig:radius_fig}
\end{figure}

To calculate the spatial coverage of the offline quotes, we borrow the concept of a ``radius of gyration'' from physical science. The radius of gyration, $R_g$, of a propaganda quote represents the typical distance travelled by the quote, weighted by number of occurrences in different locations. In our calculation, we weight each quote instance equally (the ``mass'' of each quote is 1 in the original physical science terms). 
We start with each unique quote cluster $C \in Q$, the result of our matching process described in \ref{sec:matching}.
For each set of $N$ matching propaganda quotes within a quote cluster $C = \{q_1, q_2...\}$, we calculate the location centroid $q_c$ of $C$ as the average latitude/longitude of the quotes in $C$. 
$R_g$ is the root mean square Euclidean distance of $q_i$ to the centroid $q_c$.  
Formally, $R_g$ is represented as \autoref{eq:radius} and illustrated in \autoref{fig:radius_fig}.

\begin{center}
\begin{align}
R_g &= \sqrt{ \frac{\sum_{i=1}^N (q_i-q_C)^2}{N} }, \label{eq:radius} \\
\text{where } (q_i-q_C) &= \text{Euclidean distance of } q_i \text{ to } q_C, \nonumber\\ 
q_C &= \text{centroid of quotes,} \nonumber\\ 
N &= \text{num matching quotes} \nonumber
\end{align}
\textbf{Equation 1: Calculation of Radius of Gyration}
\end{center}

\subsection{Temporal Coverage}
We analyze at what points in time offline events and the same propaganda quotes occur in online posts.
In this manner, we can observe the temporal trends per medium and analyze the temporal trends across each other.

We first calculate how long the quotes appear in reported offline events across the US, simply the number of days from the first occurrence to the final occurrence.

To get an idea of how long white supremacist propaganda quotes circulate in their first environment before crossing over to the second environment, for each quote cluster $C$ of matching quotes, we examine the temporal relationship between first appearances online and offline. We calculate the average time difference in terms of days between its first appearance in the online dataset and its first appearance in the offline dataset, or vice versa.

\section{Results}

\subsection{Geographic Distribution of Offline Propaganda Quotes}
\begin{figure}[h!]
\centering
\includegraphics[width=0.5\textwidth]{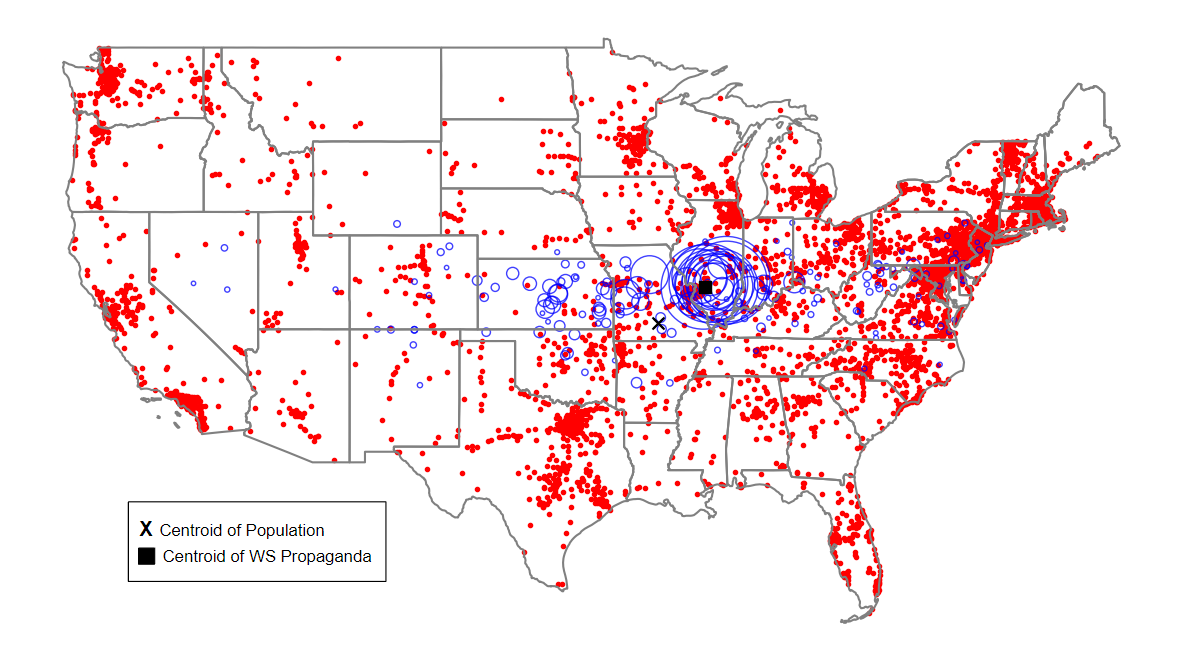}
\caption{Geographic distribution of propaganda quotes from the offline dataset. Each red dot is an incident. Each blue circle is centered around the average geographical locations of a quote, its size is proportional to the quote's frequency.}
\label{fig:usamap}
\end{figure}

\autoref{fig:usamap} depicts the geographical distribution of white supremacist propaganda quotes as extracted from the offline dataset. The blue circles are the centers of gyration for each quote cluster, sized proportionally by the number of occurrences of the quote. 
This propaganda is spread widely across the United States.
To test how the distribution of propaganda compares with the overall population distribution of the United States, we first calculate the overall center of propaganda as the average latitude and longitude across all centers of quotes, weighted by quote frequency.
If propaganda is evenly distributed across the US population, this center would match the mean center of population, the mean latitude and longitude of all population in the United States.
We find that the center of propaganda is further to the northeast than this mean center of population, which suggests that propaganda has a slight tendency toward the northeast section of the US.

\subsection{Spatiotemporal Distribution of Quotes}

\begin{figure}[h!]
\centering
\includegraphics[width=0.5\textwidth]{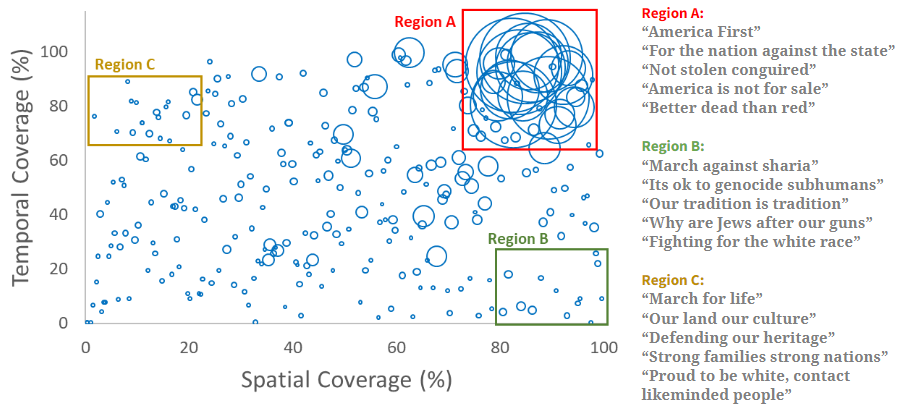}
\caption{Spatiotemporal distribution of quotes, with the most popular quotes of each region listed to the right.}
\label{fig:spatiotemporal}
\end{figure}

The spatiotemporal distribution of the offline propaganda quotes is visualized in \autoref{fig:spatiotemporal}. In Region A, quotes have both high spatial and temporal coverage. These quotes, such as ``America First'', ``America is not for sale'' are repeated extremely often and over a wide spread of area. Region B are quotes that cover a wide spread of geography but are not long-lasting (i.e., ``March against sharia''). Region C depicts quotes that are long-lasting but do not cover a large area (i.e., ``Defending our heritage''). 

We observe that propaganda messages with high spatiotemporal coverage (i.e., Region A at the top right corner of the graph) are quotable, memorable, and could apply in many contexts. 
These properties of successful propaganda are in line with the findings about the discourse of propaganda in the Persian Gulf War \cite{oddo_discourse_2019}. 
We find that the most commonly repeated propaganda are also appeals to patriotism.
An example of such a message is: 
\begin{quote}
    \textbf{Message:} ``One nation, against invasion'' \\
    \textbf{Event description:}  Patriot Front, a white supremacist group, distributed propanganda flyers at Northwest Arkansas Community College that read ``America is not for sale'', ``Reclaim America'', and ``One nation, against invasion'' \\ 
    \textbf{Frequency of quote:} 942 \\ 
    \textbf{Life span:} 820 days
\end{quote}

Looking at patterns in other regions of the spatiotemporal distribution, we observe much propaganda is time and location specific.

Some propaganda quotes are associated with the locations that they are encompassed in. Some quotes are associated with a small radius, possibly because the narrative theme within the message is fairly targeted towards the geographical community. For example, ``which way western man'' has the smallest radius of 2.6, centered in Philadelphia, Pennsylvania. This is a catchphrase that is misquoted from the 1978 book by William Simon and is mostly used by the European Heritage Association. William E. Simon, a businessman and philanthropist and the Secretary of the Treasury in 1974, was born in the city of Philadelphia. The use of the catchphrase around his birthplace could be a deliberate attempt to appeal in a local area. The quote appeared in 47 online posts and 26 offline events.

Another such message is ``Justice for Cannon Hinnant'', which has a small radius of gyration covering North Carolina to Virginia. The shooting of five year old Cannon Hinnant took place in August 2020 in Wilson, North Carolina \cite{shammas_2020}. Hinnant was a white American who was shot at point-blank by his Black neighbor, Darius Sessoms. The limited radius of the propaganda quote could be an attempt to arouse white supremacist feelings around the area of the incident by appealing to the commonality of geography. This specific quote appeared offline 40 times and was observed in 9 online posts.

Some narratives appear at a specific time, in tandem with world events. The narrative ``stop coronavirus, deport illegal aliens, close borders, stop immigration'' was first recorded on 21 March 2020 until 17 May 2020, coinciding with the beginning of the worldwide coronavirus pandemic. Another narrative rooted in conspiracy theories is ``every single aspect of the covid agenda is jewish''. This narrative arose on 18 September 2021 until 31 March 2022 as part of an effort to link anti-government sentiment (``the COVID agenda'') with antisemitic extremist views \cite{garner_2022}. 
This national and local targeting warrants further investigation into white supremacist communication tactics.

\begin{figure}[h!]
\centering
\includegraphics[width=0.5\textwidth]{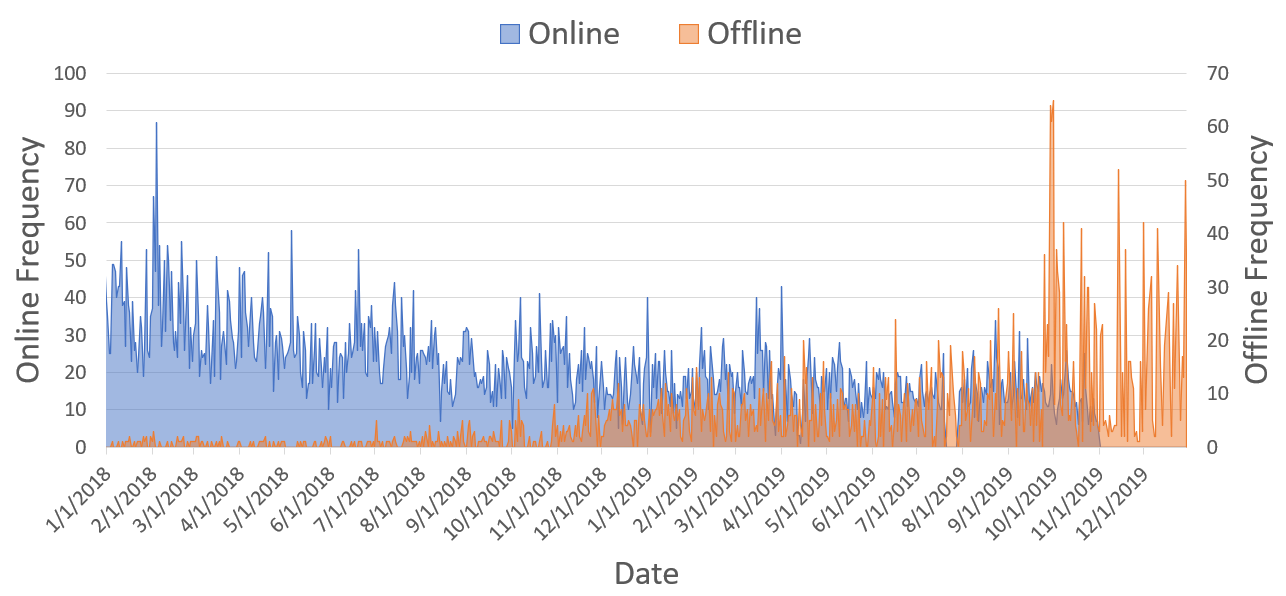}
\caption{Frequencies of propaganda quote appearances offline and online.}
\label{fig:offline_leads_online}
\end{figure}

\autoref{fig:offline_leads_online} illustrates the temporal distribution of white supremacist quotes online and offline.
This is for the 2-year period from 2018 through 2019, chosen for the presence of significant activity in both online and offline datasets.
The online dataset does not contain data past 2019, and there are relatively few reports in the offline dataset before 2018.
Overall, we find that white supremacist propaganda online leads appearances offline. 87\% of white supremacist quotes first appear online before appearing offline. The average time difference between the appearance of quotes across both medium are 242 $\pm$ 212 days. That is, quotes circulate online for an average of 7 months before making it to the offline space. Examples of such quotes are: ``white power world wide'', ``troops to the border'', ``build the wall deport them all'', ``our race is our nation'', ``only two genders''. 

\begin{figure}[h!]
\centering
\includegraphics[width=0.5\textwidth]{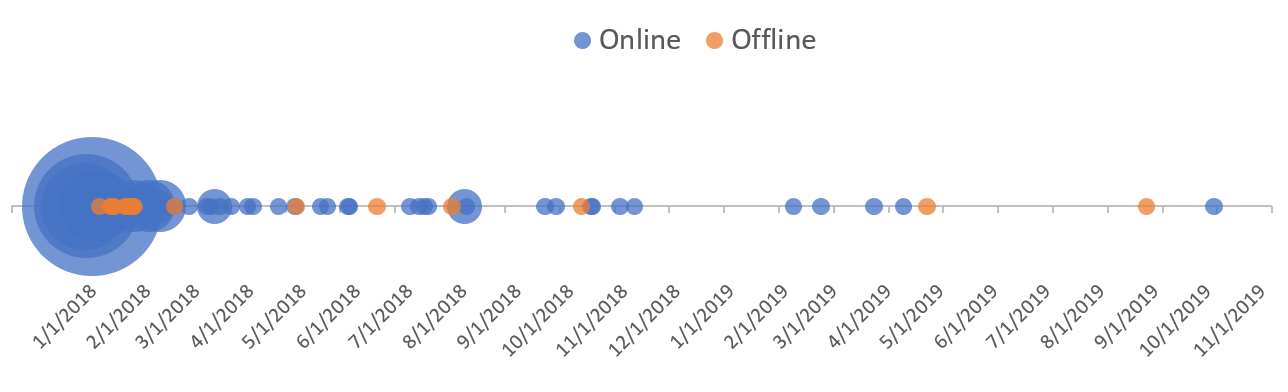}
\caption{Visualization of the first appearances of white supremacist propaganda quotes: online or offline. The size of each bubble indicates the number of unique quotes appearing on the same day. It can be observed that the vast majority of quotes first appear online}
\label{fig:offline_leads_online_2}
\end{figure}

13\% of the quotes appear offline first before making their appearance online.
Examples of such quotes are: ``protect your heritage", ``hate speech is free speech", ``wake up white America", ``Reclaim America", ``the holocaust was a good thing" . 
\autoref{fig:offline_leads_online_2} plots the first appearances of each quote in offline and online spaces.
These quotes also appear within less than two months (49 $\pm$ 158 days) online after their first offline occurrence, a much shorter period for crossover than those that appear online first.

\subsection{Comparison of Online and Offline Propaganda Use}
\begin{figure}[h!]
\centering
\includegraphics[width=0.5\textwidth]{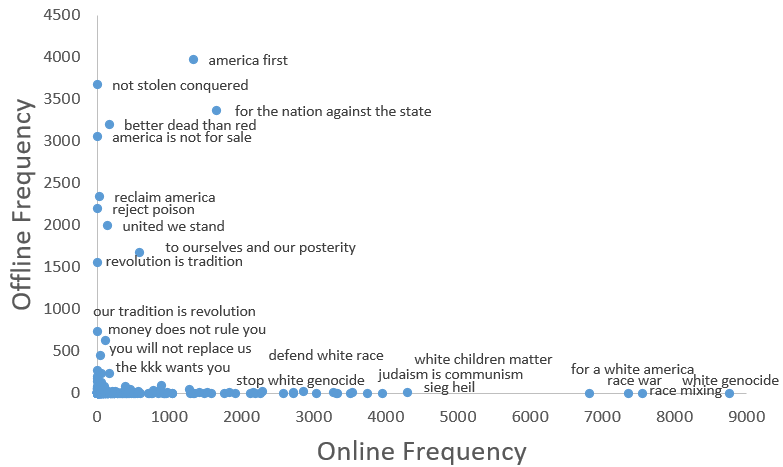}
\caption{Frequency of most popular quotes in offline and online domains.}
\label{fig:themes}
\end{figure}

We perform a comparison of propaganda use in online and offline platforms through identifying popular quotes in both environments. 
\autoref{fig:themes} compares the most popular quotes in both online and offline platforms. We observe that the most popular quotes, measured by the frequency of quote occurrences, differ between online and offline environments. 
In fact, propaganda that is popular online is not popular offline, and vice versa.

Propaganda popular in offline spaces are more vague and applicable in many contexts.
They are more likely to give plausible deniability of racism and some seem to act as dogwhistles \cite{henderson_how_2018}.
Propaganda popular online are more directly ideological, naming minoritized groups such as Jews and specifying that they are for a ``white America'' and ``white children''.

\section{Discussion}
In this study, we examined the spatiotemporal prevalence of white supremacist propaganda in both online and offline spaces. 

We find that short, decontextualized, and easily repeatable phrases like ``America First'' have the largest spatial and temporal coverage offline. 
They often appeal to patriotism and are more coded than direct.
Though white supremacist propaganda is prevalent throughout the US, we identify a trend toward the northeast area. 
There could be many reasons for this.
This area could have a historical reputation for liberalism, and so radical right-wing hate groups may view this as a ``battleground'' where propaganda is needed in a way that is not in more conservative areas.
One of the most prominent groups behind events in our dataset, contributing 84\% of the offline propaganda quotes, Patriot Front, a vastly spread hate group across the United States and is known to be active in eastern cities such as Boston and Philadelphia \cite{peterson_martin_hayden_2022}.
Some of their slogans call for the formation of a white ethno-state and restoring America to the past, which may explain a focus on this region of the country.

In the online space, we examined posts from a wide variety of data sources, ranging from online chat logs to forums to blogs to social media platforms.
This broadens our analysis beyond the idiosyncrasies of any particular platform. 
The presence of white supremacist propaganda across all these different genres of online platforms highlights the severity and spread of white supremacist ideology.

We observe white supremacist propaganda crossing from online environments to offline events.
This is in line with other online-to-offline studies, where online activism facilitates offline protests \cite{greijdanus2020psychology}. 
This points to the importance of monitoring narratives online from hate groups.

A large majority (87\%) of the quotes appear online first before appearing offline, during which they circulate for an average of seven months. 
This is a fairly long period of time, during which narratives can coalesce around certain messages before they are publicly distributed as offline propaganda.
Quotes that appear offline before we observe them in our online dataset cross over much faster, less than two months on average. 
This could be due to the ease of quickly picking up and discussing narratives in offline propaganda in online spaces (e.g. blogs, social media).

Our observations further show that propaganda that is popular in the online space do not necessarily equate to popularity in the offline space. This indicates that both environments have separate information consumption spheres, with different communicative strategies from groups with the same ideological objectives. 
Such a case has been observed in the US Capitol Riots event, an event with significant white supremacist propaganda, where there were different disinformation narratives propagated across YouTube and website videos \cite{ng2022cross}.

\subsection{Limitations and Future Work}
This study of online-to-offline crossover is not without limitations. 
Though the ADL draws from a team of expert annotators to curate its offline hate group event dataset, from which we extract our offline propaganda dataset, the annotation methodology is largely unknown. The coding scheme and the inter-annotator agreement could affect the categorization of events. 
Our online dataset does not contain posts past 2019, which limits our ability to compare with offline events.
Collecting newer online white supremacist posts (perhaps from sources such as Telegram) would allow a more recent comparison. In both temporal and spatial analysis, more stable results could be found by better-handling outliers in the dataset. 

We match propaganda quotes with literal exact matches and matches without punctuation and stopwords.
These methods could be expanded to include other forms of near matches, such as topic modeling, segmentation \cite{leskovec2009meme}, \textit{n}-gram overlap \cite{So2019}, fuzzy search \cite{bast_efficient_2013}, paraphrasing \cite{yan_are_2022}, or rhetorical function matching \cite{lin_dynamics_2020}.
These more sophisticated matching approaches may yield quotes that are similar in semantics or rhetoric, but do not have exact word overlap.

In addition, the current analysis is limited to the United States and to English propaganda. White supremacist ideology circulates worldwide, and future work should expand this analysis to other countries and contexts where messaging may vary.

An analysis of the network structure of those engaging with white supremacist propaganda offline and online may give insight to how this content is spread, and through what groups narratives cross from online to offline spaces.

\section{Conclusion}
In this work, we conduct analysis of the relationship between white supremacist propaganda in online and offline spaces. 
Understanding when and where white supremacist hate messages spread is critical for an effective and knowledgeable response by authorities, organizations, companies and academics. 

We combine data from a hand-annotated dataset that records incidents of offline white supremacist propaganda spread with a curated dataset of white supremacist quotes from online blogs, forums and social media. We begin by analyzing the geographic spread of offline propaganda, identifying that incidents of white supremacist activity are generally concentrated around regions of historical significance in the United States. 
Through temporal analysis comparison of quotes from both online and offline environments, we find preliminary evidence that online white supremacy propaganda generally appears online before being distributed offline. 
Through text comparison methods, we observe that the most popular propaganda in online and offline environments differ, but narratives seem to be time and location-specific.
We hope that this study illuminates patterns in strategic communication used by white supremacist hate groups.

\begin{acks}
The authors would like to acknowledge the support from the AFOSR awards and the Collaboratory Against Hate Research and Action Center (CAH). 
 We thank the ADL Center on Extremism for curating and providing the offline dataset and for inspiration for this project.
Any opinions, findings, and conclusions or recommendations expressed in this material do not necessarily reflect the views of the funding sources.
\end{acks}

\bibliographystyle{ACM-Reference-Format}
\bibliography{references}


\begin{thebibliography}{61}


\ifx \showCODEN    \undefined \def \showCODEN     #1{\unskip}     \fi
\ifx \showDOI      \undefined \def \showDOI       #1{#1}\fi
\ifx \showISBNx    \undefined \def \showISBNx     #1{\unskip}     \fi
\ifx \showISBNxiii \undefined \def \showISBNxiii  #1{\unskip}     \fi
\ifx \showISSN     \undefined \def \showISSN      #1{\unskip}     \fi
\ifx \showLCCN     \undefined \def \showLCCN      #1{\unskip}     \fi
\ifx \shownote     \undefined \def \shownote      #1{#1}          \fi
\ifx \showarticletitle \undefined \def \showarticletitle #1{#1}   \fi
\ifx \showURL      \undefined \def \showURL       {\relax}        \fi
\providecommand\bibfield[2]{#2}
\providecommand\bibinfo[2]{#2}
\providecommand\natexlab[1]{#1}
\providecommand\showeprint[2][]{arXiv:#2}

\bibitem[Alatawi et~al\mbox{.}(2021)]%
        {alatawi_detecting_2021}
\bibfield{author}{\bibinfo{person}{Hind~S. Alatawi}, \bibinfo{person}{Areej~M.
  Alhothali}, {and} \bibinfo{person}{Kawthar~M. Moria}.}
  \bibinfo{year}{2021}\natexlab{}.
\newblock \showarticletitle{Detecting {White} {Supremacist} {Hate} {Speech}
  {Using} {Domain} {Specific} {Word} {Embedding} with {Deep} {Learning} and
  {BERT}}.
\newblock \bibinfo{journal}{\emph{IEEE Access}}  \bibinfo{volume}{9}
  (\bibinfo{year}{2021}), \bibinfo{pages}{106363--106374}.
\newblock
\showISSN{21693536}
\urldef\tempurl%
\url{https://doi.org/10.1109/ACCESS.2021.3100435}
\showDOI{\tempurl}


\bibitem[Alorainy et~al\mbox{.}(2019)]%
        {alorainy_enemy_2019}
\bibfield{author}{\bibinfo{person}{Wafa Alorainy}, \bibinfo{person}{Pete
  Burnap}, \bibinfo{person}{Han Liu}, {and} \bibinfo{person}{Matthew~L.
  Williams}.} \bibinfo{year}{2019}\natexlab{}.
\newblock \showarticletitle{"{The} {Enemy} {Among} {Us}": {Detecting} {Cyber}
  {Hate} {Speech} with {Threats}-based {Othering} {Language} {Embeddings}}.
\newblock \bibinfo{journal}{\emph{ACM Transactions on the Web}}
  \bibinfo{volume}{13}, \bibinfo{number}{3} (\bibinfo{date}{July}
  \bibinfo{year}{2019}), \bibinfo{pages}{14:1--14:26}.
\newblock
\showISSN{1559-1131}
\urldef\tempurl%
\url{https://doi.org/10.1145/3324997}
\showDOI{\tempurl}


\bibitem[Ansah(2021)]%
        {ansah_violent_2021}
\bibfield{author}{\bibinfo{person}{Tawia Ansah}.}
  \bibinfo{year}{2021}\natexlab{}.
\newblock \showarticletitle{Violent words: strategies and legal impacts of
  white supremacist language}.
\newblock \bibinfo{journal}{\emph{Virginia Journal of Social Policy \& the
  Law}} \bibinfo{volume}{28}, \bibinfo{number}{3} (\bibinfo{year}{2021}),
  \bibinfo{pages}{305--340}.
\newblock


\bibitem[Ansley(1989)]%
        {ansley_stirring_1989}
\bibfield{author}{\bibinfo{person}{Frances~Lee Ansley}.}
  \bibinfo{year}{1989}\natexlab{}.
\newblock \showarticletitle{Stirring the {Ashes}: {Race} {Class} and the
  {Future} of {Civil} {Rights} {Scholarship}}.
\newblock \bibinfo{journal}{\emph{Cornell Law Review}} \bibinfo{volume}{74},
  \bibinfo{number}{6} (\bibinfo{year}{1989}), \bibinfo{pages}{993--1077}.
\newblock
\urldef\tempurl%
\url{http://scholarship.law.cornell.edu/clrhttp://scholarship.law.cornell.edu/clr/vol74/iss6/1}
\showURL{%
\tempurl}


\bibitem[Arnaudo(2017)]%
        {arnaudo2017computational}
\bibfield{author}{\bibinfo{person}{Dan Arnaudo}.}
  \bibinfo{year}{2017}\natexlab{}.
\newblock \showarticletitle{Computational propaganda in Brazil: Social bots
  during elections}.
\newblock  (\bibinfo{year}{2017}).
\newblock


\bibitem[Barr{\'o}n-Cedeno et~al\mbox{.}(2019)]%
        {barron2019proppy}
\bibfield{author}{\bibinfo{person}{Alberto Barr{\'o}n-Cedeno},
  \bibinfo{person}{Israa Jaradat}, \bibinfo{person}{Giovanni Da~San~Martino},
  {and} \bibinfo{person}{Preslav Nakov}.} \bibinfo{year}{2019}\natexlab{}.
\newblock \showarticletitle{Proppy: Organizing the news based on their
  propagandistic content}.
\newblock \bibinfo{journal}{\emph{Information Processing \& Management}}
  \bibinfo{volume}{56}, \bibinfo{number}{5} (\bibinfo{year}{2019}),
  \bibinfo{pages}{1849--1864}.
\newblock


\bibitem[Bast and Celikik(2013)]%
        {bast_efficient_2013}
\bibfield{author}{\bibinfo{person}{Hannah Bast} {and} \bibinfo{person}{Marjan
  Celikik}.} \bibinfo{year}{2013}\natexlab{}.
\newblock \showarticletitle{Efficient fuzzy search in large text collections}.
\newblock \bibinfo{journal}{\emph{ACM Transactions on Information Systems}}
  \bibinfo{volume}{31}, \bibinfo{number}{2} (\bibinfo{date}{May}
  \bibinfo{year}{2013}), \bibinfo{pages}{10:1--10:59}.
\newblock
\showISSN{1046-8188}
\urldef\tempurl%
\url{https://doi.org/10.1145/2457465.2457470}
\showDOI{\tempurl}


\bibitem[Berlet and Vysotsky(2006)]%
        {berlet_overview_2006}
\bibfield{author}{\bibinfo{person}{Chip Berlet} {and}
  \bibinfo{person}{Stanislav Vysotsky}.} \bibinfo{year}{2006}\natexlab{}.
\newblock \showarticletitle{Overview of {U}.{S}. {White} {Supremacist}
  {Groups}}.
\newblock \bibinfo{journal}{\emph{Journal of Political and Military Sociology}}
  \bibinfo{volume}{34}, \bibinfo{number}{1} (\bibinfo{year}{2006}),
  \bibinfo{pages}{11--48}.
\newblock


\bibitem[Blout and Burkart(2020)]%
        {blout2020white}
\bibfield{author}{\bibinfo{person}{Emily Blout} {and} \bibinfo{person}{Patrick
  Burkart}.} \bibinfo{year}{2020}\natexlab{}.
\newblock \showarticletitle{White Supremacist Terrorism in Charlottesville:
  Reconstructing ‘Unite the Right’}.
\newblock \bibinfo{journal}{\emph{Studies in Conflict \& Terrorism}}
  (\bibinfo{year}{2020}), \bibinfo{pages}{1--22}.
\newblock


\bibitem[Brown(2009)]%
        {brown_wwwhatecom_2009}
\bibfield{author}{\bibinfo{person}{Christopher Brown}.}
  \bibinfo{year}{2009}\natexlab{}.
\newblock \showarticletitle{{WWW}.{HATE}.{COM}: {White} supremacist discourse
  on the internet and the construction of whiteness ideology}.
\newblock \bibinfo{journal}{\emph{Howard Journal of Communications}}
  \bibinfo{volume}{20}, \bibinfo{number}{2} (\bibinfo{date}{April}
  \bibinfo{year}{2009}), \bibinfo{pages}{189--208}.
\newblock
\showISSN{10646175}
\urldef\tempurl%
\url{https://doi.org/10.1080/10646170902869544}
\showDOI{\tempurl}


\bibitem[Calderón et~al\mbox{.}(2021)]%
        {calderon_linguistic_2021}
\bibfield{author}{\bibinfo{person}{Fernando~H. Calderón},
  \bibinfo{person}{Namrita Balani}, \bibinfo{person}{Jherez Taylor},
  \bibinfo{person}{Melvyn Peignon}, \bibinfo{person}{Yen-Hao Huang}, {and}
  \bibinfo{person}{Yi-Shin Chen}.} \bibinfo{year}{2021}\natexlab{}.
\newblock \showarticletitle{Linguistic {Patterns} for {Code} {Word} {Resilient}
  {Hate} {Speech} {Identification}}.
\newblock \bibinfo{journal}{\emph{Sensors}} \bibinfo{volume}{21},
  \bibinfo{number}{23} (\bibinfo{date}{Jan.} \bibinfo{year}{2021}),
  \bibinfo{pages}{7859}.
\newblock
\showISSN{1424-8220}
\urldef\tempurl%
\url{https://doi.org/10.3390/s21237859}
\showDOI{\tempurl}


\bibitem[Caswell(2021)]%
        {caswell_urgent_2021}
\bibfield{author}{\bibinfo{person}{Michelle Caswell}.}
  \bibinfo{year}{2021}\natexlab{}.
\newblock \bibinfo{booktitle}{\emph{Urgent {Archives}: {Enacting} {Liberatory}
  {Memory} {Work}}}.
\newblock \bibinfo{publisher}{Taylor \& Francis}.
\newblock
\showISBNx{978-1-00-038606-6}
\showLCCN{2020053440}
\urldef\tempurl%
\url{https://books.google.com/books?id=76AsEAAAQBAJ}
\showURL{%
\tempurl}


\bibitem[Chetty and Alathur(2018)]%
        {chetty2018hate}
\bibfield{author}{\bibinfo{person}{Naganna Chetty} {and}
  \bibinfo{person}{Sreejith Alathur}.} \bibinfo{year}{2018}\natexlab{}.
\newblock \showarticletitle{Hate speech review in the context of online social
  networks}.
\newblock \bibinfo{journal}{\emph{Aggression and violent behavior}}
  \bibinfo{volume}{40} (\bibinfo{year}{2018}), \bibinfo{pages}{108--118}.
\newblock


\bibitem[Cohen et~al\mbox{.}(2014)]%
        {cohen2014detecting}
\bibfield{author}{\bibinfo{person}{Katie Cohen}, \bibinfo{person}{Fredrik
  Johansson}, \bibinfo{person}{Lisa Kaati}, {and}
  \bibinfo{person}{Jonas~Clausen Mork}.} \bibinfo{year}{2014}\natexlab{}.
\newblock \showarticletitle{Detecting linguistic markers for radical violence
  in social media}.
\newblock \bibinfo{journal}{\emph{Terrorism and Political Violence}}
  \bibinfo{volume}{26}, \bibinfo{number}{1} (\bibinfo{year}{2014}),
  \bibinfo{pages}{246--256}.
\newblock


\bibitem[Da~San~Martino et~al\mbox{.}(2021)]%
        {da2021survey}
\bibfield{author}{\bibinfo{person}{Giovanni Da~San~Martino},
  \bibinfo{person}{Stefano Cresci}, \bibinfo{person}{Alberto
  Barr{\'o}n-Cede{\~n}o}, \bibinfo{person}{Seunghak Yu},
  \bibinfo{person}{Roberto Di~Pietro}, {and} \bibinfo{person}{Preslav Nakov}.}
  \bibinfo{year}{2021}\natexlab{}.
\newblock \showarticletitle{A survey on computational propaganda detection}. In
  \bibinfo{booktitle}{\emph{Proceedings of the Twenty-Ninth International
  Conference on International Joint Conferences on Artificial Intelligence}}.
  \bibinfo{pages}{4826--4832}.
\newblock


\bibitem[Da~San~Martino et~al\mbox{.}(2019)]%
        {da-san-martino-etal-2019-fine}
\bibfield{author}{\bibinfo{person}{Giovanni Da~San~Martino},
  \bibinfo{person}{Seunghak Yu}, \bibinfo{person}{Alberto
  Barr{\'o}n-Cede{\~n}o}, \bibinfo{person}{Rostislav Petrov}, {and}
  \bibinfo{person}{Preslav Nakov}.} \bibinfo{year}{2019}\natexlab{}.
\newblock \showarticletitle{Fine-Grained Analysis of Propaganda in News
  Article}. In \bibinfo{booktitle}{\emph{Proceedings of the 2019 Conference on
  Empirical Methods in Natural Language Processing and the 9th International
  Joint Conference on Natural Language Processing (EMNLP-IJCNLP)}}.
  \bibinfo{publisher}{Association for Computational Linguistics},
  \bibinfo{address}{Hong Kong, China}, \bibinfo{pages}{5636--5646}.
\newblock
\urldef\tempurl%
\url{https://doi.org/10.18653/v1/D19-1565}
\showDOI{\tempurl}


\bibitem[Daniels(2009)]%
        {daniels2009cyber}
\bibfield{author}{\bibinfo{person}{Jessie Daniels}.}
  \bibinfo{year}{2009}\natexlab{}.
\newblock \bibinfo{booktitle}{\emph{Cyber racism: White supremacy online and
  the new attack on civil rights}}.
\newblock \bibinfo{publisher}{Rowman \& Littlefield Publishers}.
\newblock


\bibitem[Eddington(2018)]%
        {eddington2018communicative}
\bibfield{author}{\bibinfo{person}{Sean~M Eddington}.}
  \bibinfo{year}{2018}\natexlab{}.
\newblock \showarticletitle{The communicative constitution of hate
  organizations online: A semantic network analysis of “Make America Great
  Again”}.
\newblock \bibinfo{journal}{\emph{Social Media+ Society}} \bibinfo{volume}{4},
  \bibinfo{number}{3} (\bibinfo{year}{2018}),
  \bibinfo{pages}{2056305118790763}.
\newblock


\bibitem[ElSherief et~al\mbox{.}(2021)]%
        {elsherief-etal-2021-latent}
\bibfield{author}{\bibinfo{person}{Mai ElSherief}, \bibinfo{person}{Caleb
  Ziems}, \bibinfo{person}{David Muchlinski}, \bibinfo{person}{Vaishnavi
  Anupindi}, \bibinfo{person}{Jordyn Seybolt}, \bibinfo{person}{Munmun
  De~Choudhury}, {and} \bibinfo{person}{Diyi Yang}.}
  \bibinfo{year}{2021}\natexlab{}.
\newblock \showarticletitle{Latent Hatred: A Benchmark for Understanding
  Implicit Hate Speech}. In \bibinfo{booktitle}{\emph{Proceedings of the 2021
  Conference on Empirical Methods in Natural Language Processing}}.
  \bibinfo{publisher}{Association for Computational Linguistics},
  \bibinfo{address}{Online and Punta Cana, Dominican Republic},
  \bibinfo{pages}{345--363}.
\newblock
\urldef\tempurl%
\url{https://doi.org/10.18653/v1/2021.emnlp-main.29}
\showDOI{\tempurl}


\bibitem[Ferber(2004)]%
        {ferber2004home}
\bibfield{editor}{\bibinfo{person}{Abby~L. Ferber}} (Ed.).
  \bibinfo{year}{2004}\natexlab{}.
\newblock \bibinfo{booktitle}{\emph{Home-grown hate: Gender and organized
  racism}}.
\newblock \bibinfo{publisher}{Psychology Press}.
\newblock


\bibitem[Ferber and Kimmel(2000)]%
        {ferber_reading_2000}
\bibfield{author}{\bibinfo{person}{Abby~L. Ferber} {and}
  \bibinfo{person}{Michael Kimmel}.} \bibinfo{year}{2000}\natexlab{}.
\newblock \showarticletitle{Reading right: the {Western} tradition in white
  supremacist discourse}.
\newblock \bibinfo{journal}{\emph{Sociological Focus}} \bibinfo{volume}{33},
  \bibinfo{number}{2} (\bibinfo{year}{2000}), \bibinfo{pages}{193--213}.
\newblock


\bibitem[Fortuna and Nunes(2018)]%
        {fortuna2018survey}
\bibfield{author}{\bibinfo{person}{Paula Fortuna} {and}
  \bibinfo{person}{S{\'e}rgio Nunes}.} \bibinfo{year}{2018}\natexlab{}.
\newblock \showarticletitle{A survey on automatic detection of hate speech in
  text}.
\newblock \bibinfo{journal}{\emph{ACM Computing Surveys (CSUR)}}
  \bibinfo{volume}{51}, \bibinfo{number}{4} (\bibinfo{year}{2018}),
  \bibinfo{pages}{1--30}.
\newblock


\bibitem[Gagliardone et~al\mbox{.}(2015)]%
        {gagliardone2015countering}
\bibfield{author}{\bibinfo{person}{Iginio Gagliardone}, \bibinfo{person}{Danit
  Gal}, \bibinfo{person}{Thiago Alves}, {and} \bibinfo{person}{Gabriela
  Martinez}.} \bibinfo{year}{2015}\natexlab{}.
\newblock \bibinfo{booktitle}{\emph{Countering online hate speech}}.
\newblock \bibinfo{publisher}{Unesco Publishing}.
\newblock


\bibitem[Gallacher and Heerdink(2021)]%
        {gallacher_mutual_2021}
\bibfield{author}{\bibinfo{person}{John Gallacher} {and} \bibinfo{person}{Marc
  Heerdink}.} \bibinfo{year}{2021}\natexlab{}.
\newblock \bibinfo{title}{Mutual radicalisation of opposing extremist groups
  via the {Internet}}.
\newblock
\newblock
\urldef\tempurl%
\url{https://doi.org/10.31234/osf.io/dtfc5}
\showDOI{\tempurl}


\bibitem[Garner et~al\mbox{.}(2022)]%
        {garner_2022}
\bibfield{author}{\bibinfo{person}{Grace Garner}, \bibinfo{person}{Madeleine
  McGrann}, \bibinfo{person}{Maja Lynn}, \bibinfo{person}{Rachel Kranson},
  \bibinfo{person}{Michael~Miller Yoder}, {and} \bibinfo{person}{Daniel Klug}.}
  \bibinfo{year}{2022}\natexlab{}.
\newblock \bibinfo{booktitle}{\emph{The Relationship Between Antisemitism and
  COVID-19 Conspiracy on Twitter}}.
\newblock \bibinfo{type}{{T}echnical {R}eport}.
  \bibinfo{institution}{Collaboratory Against Hate}.
\newblock
\urldef\tempurl%
\url{https://www.collabagainsthate.org/papers-presentations/antisemitism-and-covid-19-conspiracy-on-twitter}
\showURL{%
\tempurl}


\bibitem[Greijdanus et~al\mbox{.}(2020)]%
        {greijdanus2020psychology}
\bibfield{author}{\bibinfo{person}{Hedy Greijdanus}, \bibinfo{person}{Carlos~A
  de Matos~Fernandes}, \bibinfo{person}{Felicity Turner-Zwinkels},
  \bibinfo{person}{Ali Honari}, \bibinfo{person}{Carla~A Roos},
  \bibinfo{person}{Hannes Rosenbusch}, {and} \bibinfo{person}{Tom Postmes}.}
  \bibinfo{year}{2020}\natexlab{}.
\newblock \showarticletitle{The psychology of online activism and social
  movements: Relations between online and offline collective action}.
\newblock \bibinfo{journal}{\emph{Current opinion in psychology}}
  \bibinfo{volume}{35} (\bibinfo{year}{2020}), \bibinfo{pages}{49--54}.
\newblock


\bibitem[Guhl and Davey(2020)]%
        {guhl2020safe}
\bibfield{author}{\bibinfo{person}{Jakob Guhl} {and} \bibinfo{person}{Jacob
  Davey}.} \bibinfo{year}{2020}\natexlab{}.
\newblock \showarticletitle{A safe space to hate: White supremacist
  mobilisation on Telegram}.
\newblock \bibinfo{journal}{\emph{Institute for Strategic Dialogue}}
  \bibinfo{volume}{26} (\bibinfo{year}{2020}).
\newblock


\bibitem[Henderson and McCready(2018)]%
        {henderson_how_2018}
\bibfield{author}{\bibinfo{person}{R. Henderson} {and} \bibinfo{person}{Elin
  McCready}.} \bibinfo{year}{2018}\natexlab{}.
\newblock \showarticletitle{How {Dogwhistles} {Work}}. In
  \bibinfo{booktitle}{\emph{New {Frontiers} in {Artificial} {Intelligence}}}
  \emph{(\bibinfo{series}{Lecture {Notes} in {Computer} {Science}})},
  \bibfield{editor}{\bibinfo{person}{Sachiyo Arai}, \bibinfo{person}{Kazuhiro
  Kojima}, \bibinfo{person}{Koji Mineshima}, \bibinfo{person}{Daisuke Bekki},
  \bibinfo{person}{Ken Satoh}, {and} \bibinfo{person}{Yuiko Ohta}} (Eds.).
  \bibinfo{publisher}{Springer International Publishing},
  \bibinfo{address}{Cham}, \bibinfo{pages}{231--240}.
\newblock
\showISBNx{978-3-319-93794-6}
\urldef\tempurl%
\url{https://doi.org/10.1007/978-3-319-93794-6_16}
\showDOI{\tempurl}


\bibitem[Hirvonen(2013)]%
        {hirvonen2013sweden}
\bibfield{author}{\bibinfo{person}{Katrina Hirvonen}.}
  \bibinfo{year}{2013}\natexlab{}.
\newblock \showarticletitle{Sweden: When hate becomes the norm}.
\newblock \bibinfo{journal}{\emph{Race \& Class}} \bibinfo{volume}{55},
  \bibinfo{number}{1} (\bibinfo{year}{2013}), \bibinfo{pages}{78--86}.
\newblock


\bibitem[Howard and Kollanyi(2016)]%
        {howard2016bots}
\bibfield{author}{\bibinfo{person}{Philip~N Howard} {and}
  \bibinfo{person}{Bence Kollanyi}.} \bibinfo{year}{2016}\natexlab{}.
\newblock \showarticletitle{Bots,\# StrongerIn, and\# Brexit: computational
  propaganda during the UK-EU referendum}.
\newblock \bibinfo{journal}{\emph{arXiv preprint arXiv:1606.06356}}
  (\bibinfo{year}{2016}).
\newblock


\bibitem[Hutchinson et~al\mbox{.}(2022)]%
        {hutchinson_online_2022}
\bibfield{author}{\bibinfo{person}{Jade Hutchinson}, \bibinfo{person}{Muhammad
  Iqbal}, \bibinfo{person}{Mario Peucker}, {and} \bibinfo{person}{Debra
  Smith}.} \bibinfo{year}{2022}\natexlab{}.
\newblock \showarticletitle{Online and {Offline} {Coordination} in
  {Australia}’s {Far}-{Right}: {A} {Study} of {True} {Blue} {Crew}}.
\newblock \bibinfo{journal}{\emph{Social Sciences}} \bibinfo{volume}{11},
  \bibinfo{number}{9} (\bibinfo{date}{Sept.} \bibinfo{year}{2022}),
  \bibinfo{pages}{421}.
\newblock
\showISSN{2076-0760}
\urldef\tempurl%
\url{https://doi.org/10.3390/socsci11090421}
\showDOI{\tempurl}
\newblock
\shownote{Number: 9 Publisher: Multidisciplinary Digital Publishing Institute}.


\bibitem[Jokubauskaitė and Peeters(2020)]%
        {jokubauskaite_generally_2020}
\bibfield{author}{\bibinfo{person}{Emilija Jokubauskaitė} {and}
  \bibinfo{person}{Stijn Peeters}.} \bibinfo{year}{2020}\natexlab{}.
\newblock \showarticletitle{Generally {Curious}: {Thematically} {Distinct}
  {Datasets} of {General} {Threads} on 4chan/pol/}. In
  \bibinfo{booktitle}{\emph{Proceedings of the {International} {AAAI}
  {Conference} on {Web} and {Social} {Media}}}, Vol.~\bibinfo{volume}{14}.
  \bibinfo{pages}{863--867}.
\newblock
\urldef\tempurl%
\url{https://ojs.aaai.org/index.php/ICWSM/article/view/7351}
\showURL{%
\tempurl}


\bibitem[Keum et~al\mbox{.}(2022)]%
        {keum2022hate}
\bibfield{author}{\bibinfo{person}{Brian~TaeHyuk Keum}, \bibinfo{person}{Xu
  Li}, {and} \bibinfo{person}{Michele~J Wong}.}
  \bibinfo{year}{2022}\natexlab{}.
\newblock \showarticletitle{Hate as a system: Examining hate crimes and hate
  groups as state level moderators on the impact of online and offline racism
  on mental health}.
\newblock \bibinfo{journal}{\emph{International Journal of Intercultural
  Relations}}  \bibinfo{volume}{91} (\bibinfo{year}{2022}),
  \bibinfo{pages}{44--55}.
\newblock


\bibitem[LeFebvre and Armstrong(2018)]%
        {lefebvre2018grievance}
\bibfield{author}{\bibinfo{person}{Rebecca~Kay LeFebvre} {and}
  \bibinfo{person}{Crystal Armstrong}.} \bibinfo{year}{2018}\natexlab{}.
\newblock \showarticletitle{Grievance-based social movement mobilization in
  the\# Ferguson Twitter storm}.
\newblock \bibinfo{journal}{\emph{New Media \& Society}} \bibinfo{volume}{20},
  \bibinfo{number}{1} (\bibinfo{year}{2018}), \bibinfo{pages}{8--28}.
\newblock


\bibitem[Leskovec et~al\mbox{.}(2009)]%
        {leskovec2009meme}
\bibfield{author}{\bibinfo{person}{Jure Leskovec}, \bibinfo{person}{Lars
  Backstrom}, {and} \bibinfo{person}{Jon Kleinberg}.}
  \bibinfo{year}{2009}\natexlab{}.
\newblock \showarticletitle{Meme-tracking and the dynamics of the news cycle}.
  In \bibinfo{booktitle}{\emph{Proceedings of the 15th {ACM} {SIGKDD}
  international conference on {Knowledge} discovery and data mining}}.
  \bibinfo{pages}{497--506}.
\newblock
\urldef\tempurl%
\url{https://dl.acm.org/doi/pdf/10.1145/1557019.1557077?casa_token=G2JLp0oyyS8AAAAA:Q0od5ta0fvnLV-3eXclr7idBYqnJJEQ5NaLpqYjcPW40EUsdHJeX7tDgsRiXYtMxun-0B_VParTpMw}
\showURL{%
\tempurl}


\bibitem[Lin and Chung(2020)]%
        {lin_dynamics_2020}
\bibfield{author}{\bibinfo{person}{Yu-Ru Lin} {and} \bibinfo{person}{Wen-Ting
  Chung}.} \bibinfo{year}{2020}\natexlab{}.
\newblock \showarticletitle{The dynamics of {Twitter} users’ gun narratives
  across major mass shooting events}.
\newblock \bibinfo{journal}{\emph{Humanities and Social Sciences
  Communications}} \bibinfo{volume}{7}, \bibinfo{number}{1}
  (\bibinfo{date}{Aug.} \bibinfo{year}{2020}), \bibinfo{pages}{1--16}.
\newblock
\showISSN{2662-9992}
\urldef\tempurl%
\url{https://doi.org/10.1057/s41599-020-00533-8}
\showDOI{\tempurl}
\newblock
\shownote{Number: 1 Publisher: Palgrave}.


\bibitem[Lupu et~al\mbox{.}(2023)]%
        {lupu2023offline}
\bibfield{author}{\bibinfo{person}{Yonatan Lupu}, \bibinfo{person}{Richard
  Sear}, \bibinfo{person}{Nicolas Vel{\'a}squez}, \bibinfo{person}{Rhys Leahy},
  \bibinfo{person}{Nicholas~Johnson Restrepo}, \bibinfo{person}{Beth Goldberg},
  {and} \bibinfo{person}{Neil~F Johnson}.} \bibinfo{year}{2023}\natexlab{}.
\newblock \showarticletitle{Offline events and online hate}.
\newblock \bibinfo{journal}{\emph{Plos one}} \bibinfo{volume}{18},
  \bibinfo{number}{1} (\bibinfo{year}{2023}), \bibinfo{pages}{e0278511}.
\newblock


\bibitem[Nakov et~al\mbox{.}(2022)]%
        {10.1145/3488560.3501395}
\bibfield{author}{\bibinfo{person}{Preslav Nakov}, \bibinfo{person}{Giovanni
  Da~San~Martino}, {and} \bibinfo{person}{Firoj Alam}.}
  \bibinfo{year}{2022}\natexlab{}.
\newblock \showarticletitle{Fact-Checking, Fake News, Propaganda, Media Bias,
  and the COVID-19 Infodemic}. In \bibinfo{booktitle}{\emph{Proceedings of the
  Fifteenth ACM International Conference on Web Search and Data Mining}}
  (Virtual Event, AZ, USA) \emph{(\bibinfo{series}{WSDM '22})}.
  \bibinfo{publisher}{Association for Computing Machinery},
  \bibinfo{address}{New York, NY, USA}, \bibinfo{pages}{1632–1634}.
\newblock
\showISBNx{9781450391320}
\urldef\tempurl%
\url{https://doi.org/10.1145/3488560.3501395}
\showDOI{\tempurl}


\bibitem[Ng et~al\mbox{.}(2022)]%
        {ng2022cross}
\bibfield{author}{\bibinfo{person}{Lynnette Hui~Xian Ng},
  \bibinfo{person}{Iain~J Cruickshank}, {and} \bibinfo{person}{Kathleen~M
  Carley}.} \bibinfo{year}{2022}\natexlab{}.
\newblock \showarticletitle{Cross-platform information spread during the
  january 6th capitol riots}.
\newblock \bibinfo{journal}{\emph{Social Network Analysis and Mining}}
  \bibinfo{volume}{12}, \bibinfo{number}{1} (\bibinfo{year}{2022}),
  \bibinfo{pages}{133}.
\newblock


\bibitem[Ng and Loke(2020)]%
        {ng2020analyzing}
\bibfield{author}{\bibinfo{person}{Lynnette Hui~Xian Ng} {and}
  \bibinfo{person}{Jia~Yuan Loke}.} \bibinfo{year}{2020}\natexlab{}.
\newblock \showarticletitle{Analyzing public opinion and misinformation in a
  COVID-19 telegram group chat}.
\newblock \bibinfo{journal}{\emph{IEEE Internet Computing}}
  \bibinfo{volume}{25}, \bibinfo{number}{2} (\bibinfo{year}{2020}),
  \bibinfo{pages}{84--91}.
\newblock


\bibitem[Oddo(2019)]%
        {oddo_discourse_2019}
\bibfield{author}{\bibinfo{person}{John Oddo}.}
  \bibinfo{year}{2019}\natexlab{}.
\newblock \bibinfo{booktitle}{\emph{The {Discourse} of {Propaganda}: {Case}
  {Studies} from the {Persian} {Gulf} {War} and the {War} on {Terror}}}.
\newblock \bibinfo{publisher}{Penn State Press}.
\newblock
\showISBNx{978-0-271-08273-8}


\bibitem[Papasavva et~al\mbox{.}(2020)]%
        {papasavva_raiders_2020}
\bibfield{author}{\bibinfo{person}{Antonis Papasavva}, \bibinfo{person}{Savvas
  Zannettou}, \bibinfo{person}{Emiliano~De Cristofaro},
  \bibinfo{person}{Gianluca Stringhini}, {and} \bibinfo{person}{Jeremy
  Blackburn}.} \bibinfo{year}{2020}\natexlab{}.
\newblock \showarticletitle{Raiders of the {Lost} {Kek}: 3.5 {Years} of
  {Augmented} 4chan {Posts} from the {Politically} {Incorrect} {Board}}.
\newblock \bibinfo{journal}{\emph{Proceedings of the International AAAI
  Conference on Web and Social Media}}  \bibinfo{volume}{14}
  (\bibinfo{date}{May} \bibinfo{year}{2020}), \bibinfo{pages}{885--894}.
\newblock
\showISSN{2334-0770}
\urldef\tempurl%
\url{https://ojs.aaai.org/index.php/ICWSM/article/view/7354}
\showURL{%
\tempurl}


\bibitem[Perry and Scrivens(2016)]%
        {perry2016white}
\bibfield{author}{\bibinfo{person}{Barbara Perry} {and} \bibinfo{person}{Ryan
  Scrivens}.} \bibinfo{year}{2016}\natexlab{}.
\newblock \showarticletitle{White pride worldwide: Constructing global
  identities online}.
\newblock \bibinfo{journal}{\emph{The globalisation of hate: Internationalising
  hate crime}} (\bibinfo{year}{2016}), \bibinfo{pages}{65--78}.
\newblock


\bibitem[Peterson et~al\mbox{.}(2022)]%
        {peterson_martin_hayden_2022}
\bibfield{author}{\bibinfo{person}{Kevin~C Peterson}, \bibinfo{person}{Phillip
  Martin}, {and} \bibinfo{person}{Michael~Edison Hayden}.}
  \bibinfo{year}{2022}\natexlab{}.
\newblock \bibinfo{title}{"Children of the kkk": White supremacist Patriot
  Front marches through Boston, attacks Black Artist}.
\newblock
\newblock
\urldef\tempurl%
\url{https://www.democracynow.org/2022/7/6/neo_nazi_nsc_131_patriot_front}
\showURL{%
\tempurl}


\bibitem[Phillips et~al\mbox{.}(2018)]%
        {phillips2018daily}
\bibfield{author}{\bibinfo{person}{Matthew Phillips},
  \bibinfo{person}{Arunkumar Bagavathi}, \bibinfo{person}{Shannon~E Reid},
  \bibinfo{person}{Matthew Valasik}, {and} \bibinfo{person}{Siddharth
  Krishnan}.} \bibinfo{year}{2018}\natexlab{}.
\newblock \showarticletitle{The daily use of Gab is climbing. Which talker
  might become as violent as the Pittsburgh synagogue gunman}.
\newblock \bibinfo{journal}{\emph{The Washington Post}} (\bibinfo{year}{2018}).
\newblock


\bibitem[Pruden et~al\mbox{.}(2022)]%
        {pruden_birds_2022}
\bibfield{author}{\bibinfo{person}{Meredith~L. Pruden},
  \bibinfo{person}{Ayse~D. Lokmanoglu}, \bibinfo{person}{Anne Peterscheck},
  {and} \bibinfo{person}{Yannick Veilleux-Lepage}.}
  \bibinfo{year}{2022}\natexlab{}.
\newblock \showarticletitle{Birds of a {Feather}: {A} {Comparative} {Analysis}
  of {White} {Supremacist} and {Violent} {Male} {Supremacist} {Discourses}}.
\newblock In \bibinfo{booktitle}{\emph{Right-{Wing} {Extremism} in {Canada} and
  the {United} {States}}}. \bibinfo{publisher}{Palgrave Macmillan},
  \bibinfo{pages}{215--254}.
\newblock
\urldef\tempurl%
\url{https://doi.org/10.1007/978-3-030-99804-2_9}
\showDOI{\tempurl}


\bibitem[Qian et~al\mbox{.}(2018)]%
        {qian_hierarchical_2018}
\bibfield{author}{\bibinfo{person}{Jing Qian}, \bibinfo{person}{Mai Elsherief},
  \bibinfo{person}{Elizabeth Belding}, {and} \bibinfo{person}{William~Yang
  Wang}.} \bibinfo{year}{2018}\natexlab{}.
\newblock \showarticletitle{Hierarchical {CVAE} for {Fine}-{Grained} {Hate}
  {Speech} {Classification}}. In \bibinfo{booktitle}{\emph{Proceedings of the
  2018 {Conference} on {Empirical} {Methods} in {Natural} {Language}
  {Processing}}}. \bibinfo{pages}{3550--3559}.
\newblock


\bibitem[Quek(2019)]%
        {quek2019paso}
\bibfield{author}{\bibinfo{person}{Natasha Quek}.}
  \bibinfo{year}{2019}\natexlab{}.
\newblock \showarticletitle{El-Paso Shootings: Growing Threat of White
  Supremacists}.
\newblock \bibinfo{journal}{\emph{RSIS Commentary}} (\bibinfo{year}{2019}).
\newblock
\urldef\tempurl%
\url{https://dr.ntu.edu.sg/handle/10220/49921}
\showURL{%
\tempurl}


\bibitem[Rieger et~al\mbox{.}(2021)]%
        {rieger_assessing_2021}
\bibfield{author}{\bibinfo{person}{Diana Rieger}, \bibinfo{person}{Anna~Sophie
  Kümpel}, \bibinfo{person}{Maximilian Wich}, \bibinfo{person}{Toni Kiening},
  {and} \bibinfo{person}{Georg Groh}.} \bibinfo{year}{2021}\natexlab{}.
\newblock \showarticletitle{Assessing the {Extent} and {Types} of {Hate}
  {Speech} in {Fringe} {Communities}: {A} {Case} {Study} of {Alt}-{Right}
  {Communities} on 8chan, 4chan, and {Reddit}}.
\newblock \bibinfo{journal}{\emph{Social Media and Society}}
  \bibinfo{volume}{7}, \bibinfo{number}{4} (\bibinfo{year}{2021}).
\newblock
\showISSN{20563051}
\urldef\tempurl%
\url{https://doi.org/10.1177/20563051211052906}
\showDOI{\tempurl}


\bibitem[Shaar et~al\mbox{.}(2021)]%
        {shaar-etal-2021-findings}
\bibfield{author}{\bibinfo{person}{Shaden Shaar}, \bibinfo{person}{Firoj Alam},
  \bibinfo{person}{Giovanni Da~San~Martino}, \bibinfo{person}{Alex Nikolov},
  \bibinfo{person}{Wajdi Zaghouani}, \bibinfo{person}{Preslav Nakov}, {and}
  \bibinfo{person}{Anna Feldman}.} \bibinfo{year}{2021}\natexlab{}.
\newblock \showarticletitle{Findings of the {NLP}4{IF}-2021 Shared Tasks on
  Fighting the {COVID}-19 Infodemic and Censorship Detection}. In
  \bibinfo{booktitle}{\emph{Proceedings of the Fourth Workshop on NLP for
  Internet Freedom: Censorship, Disinformation, and Propaganda}}.
  \bibinfo{publisher}{Association for Computational Linguistics},
  \bibinfo{address}{Online}, \bibinfo{pages}{82--92}.
\newblock
\urldef\tempurl%
\url{https://doi.org/10.18653/v1/2021.nlp4if-1.12}
\showDOI{\tempurl}


\bibitem[Shammas(2020)]%
        {shammas_2020}
\bibfield{author}{\bibinfo{person}{Brittany Shammas}.}
  \bibinfo{year}{2020}\natexlab{}.
\newblock \bibinfo{title}{What we know about the killing of 5-year-old Cannon
  Hinnant}.
\newblock
\newblock
\urldef\tempurl%
\url{https://www.washingtonpost.com/nation/2020/08/14/cannon-hinnant-killing/}
\showURL{%
\tempurl}


\bibitem[Siegel et~al\mbox{.}(2021)]%
        {siegel_trumping_2021}
\bibfield{author}{\bibinfo{person}{Alexandra~A. Siegel},
  \bibinfo{person}{Evgenii Nikitin}, \bibinfo{person}{Pablo Barberá},
  \bibinfo{person}{Joanna Sterling}, \bibinfo{person}{Bethany Pullen},
  \bibinfo{person}{Richard Bonneau}, \bibinfo{person}{Jonathan Nagler}, {and}
  \bibinfo{person}{Joshua~A. Tucker}.} \bibinfo{year}{2021}\natexlab{}.
\newblock \showarticletitle{Trumping {Hate} on {Twitter}? {Online} {Hate}
  {Speech} in the 2016 {U}.{S}. {Election} {Campaign} and its {Aftermath}}.
\newblock \bibinfo{journal}{\emph{Quarterly Journal of Political Science}}
  \bibinfo{volume}{16} (\bibinfo{year}{2021}), \bibinfo{pages}{71--104}.
\newblock
\urldef\tempurl%
\url{https://www.nowpublishers.com/article/Details/QJPS-19045}
\showURL{%
\tempurl}


\bibitem[So et~al\mbox{.}(2019)]%
        {So2019}
\bibfield{author}{\bibinfo{person}{Richard So}, \bibinfo{person}{Hoyt Long},
  {and} \bibinfo{person}{Yuancheng Zhu}.} \bibinfo{year}{2019}\natexlab{}.
\newblock \showarticletitle{Race, {Writing}, and {Computation}: {Racial}
  {Difference} and the {US} {Novel}, 1880-2000}.
\newblock \bibinfo{journal}{\emph{Journal of Cultural Analytics}}
  (\bibinfo{year}{2019}), \bibinfo{pages}{1--30}.
\newblock
\urldef\tempurl%
\url{https://doi.org/10.22148/16.031}
\showDOI{\tempurl}


\bibitem[Somoano and McNeil-Willson(2022)]%
        {somoano2022lessons}
\bibfield{author}{\bibinfo{person}{In{\'e}s~Bola{\~n}os Somoano} {and}
  \bibinfo{person}{Richard McNeil-Willson}.} \bibinfo{year}{2022}\natexlab{}.
\newblock \bibinfo{booktitle}{\emph{Lessons From the Buffalo Shooting:
  Responses to Violent White Supremacy}}.
\newblock \bibinfo{type}{{T}echnical {R}eport}.
  \bibinfo{institution}{International Centre for Counter-Terrorism}.
\newblock


\bibitem[Sullivan and Benner(2021)]%
        {sullivan_top_2021}
\bibfield{author}{\bibinfo{person}{Eileen Sullivan} {and}
  \bibinfo{person}{Katie Benner}.} \bibinfo{year}{2021}\natexlab{}.
\newblock \showarticletitle{Top law enforcement officials say the biggest
  domestic terror threat comes from white supremacists.}
\newblock \bibinfo{journal}{\emph{The New York Times}} (\bibinfo{date}{May}
  \bibinfo{year}{2021}).
\newblock
\showISSN{0362-4331}
\urldef\tempurl%
\url{https://www.nytimes.com/2021/05/12/us/politics/domestic-terror-white-supremacists.html}
\showURL{%
\tempurl}


\bibitem[Wilkins et~al\mbox{.}(2019)]%
        {wilkins2019all}
\bibfield{author}{\bibinfo{person}{Denise~J Wilkins}, \bibinfo{person}{Andrew~G
  Livingstone}, {and} \bibinfo{person}{Mark Levine}.}
  \bibinfo{year}{2019}\natexlab{}.
\newblock \showarticletitle{All click, no action? Online action, efficacy
  perceptions, and prior experience combine to affect future collective
  action}.
\newblock \bibinfo{journal}{\emph{Computers in Human Behavior}}
  \bibinfo{volume}{91} (\bibinfo{year}{2019}), \bibinfo{pages}{97--105}.
\newblock


\bibitem[Windisch et~al\mbox{.}(2018)]%
        {windisch2018understanding}
\bibfield{author}{\bibinfo{person}{Steven Windisch}, \bibinfo{person}{Pete
  Simi}, \bibinfo{person}{Kathleen Blee}, {and} \bibinfo{person}{Matthew
  DeMichele}.} \bibinfo{year}{2018}\natexlab{}.
\newblock \showarticletitle{Understanding the micro-situational dynamics of
  white supremacist violence in the United States}.
\newblock \bibinfo{journal}{\emph{Perspectives on Terrorism}}
  \bibinfo{volume}{12}, \bibinfo{number}{6} (\bibinfo{year}{2018}),
  \bibinfo{pages}{23--37}.
\newblock


\bibitem[Wong et~al\mbox{.}(2015)]%
        {wong2015supremacy}
\bibfield{author}{\bibinfo{person}{Meghan~A Wong}, \bibinfo{person}{Richard
  Frank}, {and} \bibinfo{person}{Russell Allsup}.}
  \bibinfo{year}{2015}\natexlab{}.
\newblock \showarticletitle{The supremacy of online white supremacists--an
  analysis of online discussions by white supremacists}.
\newblock \bibinfo{journal}{\emph{Information \& Communications Technology
  Law}} \bibinfo{volume}{24}, \bibinfo{number}{1} (\bibinfo{year}{2015}),
  \bibinfo{pages}{41--73}.
\newblock


\bibitem[Woolley and Howard(2017)]%
        {woolley2017computational}
\bibfield{author}{\bibinfo{person}{Samuel~C Woolley} {and}
  \bibinfo{person}{Philip Howard}.} \bibinfo{year}{2017}\natexlab{}.
\newblock \showarticletitle{Computational propaganda worldwide: Executive
  summary}.
\newblock  (\bibinfo{year}{2017}).
\newblock


\bibitem[Yan et~al\mbox{.}(2022)]%
        {yan_are_2022}
\bibfield{author}{\bibinfo{person}{Muheng Yan}, \bibinfo{person}{Yu-Ru Lin},
  {and} \bibinfo{person}{Wen-Ting Chung}.} \bibinfo{year}{2022}\natexlab{}.
\newblock \showarticletitle{Are {Mutated} {Misinformation} {More} {Contagious}?
  {A} {Case} {Study} of {COVID}-19 {Misinformation} on {Twitter}}. In
  \bibinfo{booktitle}{\emph{14th {ACM} {Web} {Science} {Conference} 2022}}
  \emph{(\bibinfo{series}{{WebSci} '22})}. \bibinfo{publisher}{Association for
  Computing Machinery}, \bibinfo{address}{New York, NY, USA},
  \bibinfo{pages}{336--347}.
\newblock
\showISBNx{978-1-4503-9191-7}
\urldef\tempurl%
\url{https://doi.org/10.1145/3501247.3531562}
\showDOI{\tempurl}


\bibitem[Yoder et~al\mbox{.}(2023)]%
        {anonymous2023}
\bibfield{author}{\bibinfo{person}{Michael~Miller Yoder},
  \bibinfo{person}{Ahmad Diab}, \bibinfo{person}{David West~Brown}, {and}
  \bibinfo{person}{Kathleen~M. Carley}.} \bibinfo{year}{2023}\natexlab{}.
\newblock \showarticletitle{A Weakly Supervised Classifier and Dataset of White
  Supremacist Language}. \bibinfo{pages}{Under review}.
\newblock


\end{thebibliography}


\end{document}